\documentclass[dvipsnames]{article}
\usepackage{spconf,amsmath,graphicx,bm}
\usepackage{enumitem}
\usepackage[colorlinks=true, citecolor=cyan, linkcolor=BrickRed]{hyperref}
\usepackage{subfig}
\usepackage[export]{adjustbox}
\usepackage{flushend}

\usepackage{pgfplots}
\pgfplotsset{compat=1.9}
\usepgfplotslibrary{units, external}
\DeclareMathOperator{\x}{\mathbf{x}}
\DeclareMathOperator{\M}{\mathbf{S}}
\DeclareMathOperator{\y}{\mathbf{y}_\text{obs}}
\DeclareMathOperator{\z}{\mathbf{z}}
\DeclareMathOperator{\te}{\bm{\theta}}
\DeclareMathOperator{\F}{\mathbf{F}}
\DeclareMathOperator{\ani}{\mathbf{w}}
\DeclareMathOperator{\slopes}{\bm{\phi}}
\DeclareMathOperator{\Lapl}{\overrightarrow{\nabla}_{\slopes}}
\DeclareMathOperator{\eps}{\varepsilon}

\title{Anti-Aliasing Add-On for Deep Prior Seismic Data Interpolation}

\name{
\begin{tabular}{@{}c@{}}Francesco~Picetti \qquad
        Vincenzo~Lipari \qquad
        Paolo~Bestagini \qquad
        Stefano~Tubaro
        \end{tabular}}
\address{Image and Sound Processing Lab (ISPL), \\ Department of Electronics, Information and Bioengineering (DEIB),\\
Politecnico di Milano, Milan, Italy}

\begin{document}
%\ninept

\maketitle

\begin{abstract}
    Data interpolation is a fundamental step in any seismic processing workflow. Among machine learning techniques recently proposed to solve data interpolation as an inverse problem, Deep Prior paradigm aims at employing a convolutional neural network to capture priors on the data in order to regularize the inversion. 
    However, this technique lacks of reconstruction precision when interpolating highly decimated data due to the presence of aliasing.
    In this work, we propose to improve Deep Prior inversion by adding a directional Laplacian as regularization term to the problem.
    This regularizer drives the optimization towards solutions that honor the slopes estimated from the interpolated data low frequencies.
    We provide some numerical examples to showcase the methodology devised in this manuscript, showing that our results are less prone to aliasing also in presence of noisy and corrupted data.
\end{abstract}

\section{Introduction}
Economic or environmental constraints, cable feathering in marine acquisitions and the presence of dead or damaged traces result in most seismic datasets being poorly and often irregularly sampled in space. As a result, the vast majority of seismic processing and imaging workflows \cite{chang20063d, virieux2009overview, verschuur1992adaptive} typically include a preliminary interpolation step. The importance of the issue is confirmed by the vast number of methods proposed and the vitality of the relevant scientific literature.

Besides traditional processing methods such as transform representations \cite{spitz1991seismic, huang2018dreamlet, kumar2017beating},  low-rank approximation \cite{lopez2016offthegrid}, multichannel singular spectrum analysis (MSSA) \cite{nazari2017simultaneous} and its interpolated version I-MSSA \cite{carozzi2021interpolated},
many seismic interpolation methods based on Convolutional Neural Networks (CNNs) have been proposed
%\cite{mandelli2019interpolation, wang2018seismic, siahkoohi2018seismic, mikhailiuk2018deep, oliveira2018interpolating, kaur2019seismic, yoon2020seismic}.
\cite{mandelli2019interpolation, siahkoohi2018seismic, oliveira2018interpolating, yoon2020seismic}.
However, the vast majority of CNN-based methods work according to a training paradigm. In other words, the used CNN learns how to reconstruct missing traces during a training step, in which several pairs of corresponding corrupted and uncorrupted data are fed to the CNN itself.
On one hand, this approach needs a huge amount of training data. On the other hand, CNNs performances strongly depend on training examples seen during training, which may prevent the CNN to generalize and work on different kinds of data.

A different approach has been proposed interpreting the CNN architecture as a Deep Prior in the framework of inverse problems to address tasks such as interpolation, denoising or super-resolution \cite{Ulyanov_2018_CVPR}.
Through this approach, the CNN learns the inner structure of a 2D image from the corrupted data itself, without pre-training: this prevents any over-fitting issue as well as the need for training data. This strategy has been recently explored also for the task of 2D seismic data interpolation using common U-Net as Deep Prior.
In particular, Deep Priors have shown effective reconstruction performance when tackling both irregular and regular sampling \cite{liu2019deep}.
However, Deep Prior techniques are not optimally performant when facing highly aliased data.

Inspired by our previous studies \cite{kong2020eage, kong2020seg, kong2020grsl}, we propose to regularize the Deep Prior objective function with a directional Laplacian.
Specifically, data orientations are estimated from the low-end spectrum of the CNN output. Then, we build the directional Laplacian that drives the optimization towards solutions that minimizes the energy of events not coherent with estimated slopes.

The rest of the paper is structured as follows.
Section~\ref{sec:method} provides the theoretical formulation of the interpolation problem and the details of the proposed method.
Section~\ref{sec:results} contains the results achieved through our numerical simulations of three different scenarios.
Finally, Section~\ref{sec:conclusion} concludes the paper.
The complete code to run our analysis and the datasets used in this manuscript are accessible to the reader in a repository available at \url{https://bit.ly/3a2ef1t}.
%\url{https://github.com/polimi-ispl/deep_prior_interpolation}. 

\section{Methodology}\label{sec:method}
Seismic data interpolation is an ill-posed inverse problem.
In order to obtain a reasonable solution it is necessary to constrain the inversion by adding some kind of a priori information.
A standard way consists in formulating the inverse problem as finding the solution $\x^*$ that minimizes an objective function
\begin{equation}
    J\left(\x \right)=E\left(\M \x -\y\right)+R\left(\x\right),
    \label{eq:penalty}
\end{equation}
where $\M$ is a linear sampling operator, $\y$ are the acquired seismic data, $E(\cdot)$ is the data fidelity term, and $R(\cdot)$ is a regularizer function formalizing the distance of the solution from the space of solutions honoring the desired prior. 
The choice of $R(\cdot)$ is critical and usually derives from insights of human experts.

In Deep Prior paradigms, feasible solutions are modeled as the application of a CNN to a random noise realization $\z$.
The CNN acts as a parametric non linear function $f_{\te} \left(\cdot\right)$ and the problem is recast as finding the set of parameters $\te^*$ which minimizes the objective function
\begin{equation}
     J\left(\te\right) = E \left( \M f_{\te} \left( \z \right) -\y \right).
    \label{eq:DIP}
\end{equation}

In this work, we propose a modification to the objective cost function \eqref{eq:DIP}, adding a regularization term given by the Laplacian operator oriented according to the events slopes.

Since data is aliased due to sub-sampling, we need to estimate the slopes in a robust way.
To this end, we first estimate a low-pass version of the data by applying Deep Prior interpolation considering the cost function
\begin{equation}
     J\left(\te\right) = E \left( \M f_{\te} \left( \z \right) - \F \y \right),
    \label{eq:DIP_low}
\end{equation}
$\F$ being a 1D low-pass filter that acts trace-by-trace, thus removing alias effect that could badly impact on standard Deep Prior inversion.

By minimizing Eq.~\eqref{eq:DIP_low}, we estimate a low-pass version of the interpolated data.
Then, we employ the structure tensor algorithm \cite{vanvliet1995estimators} to estimate the slope angles $\slopes$ and to build the directional Laplacian operator $\Lapl$ \cite{hale2007local}. The latter is employed to regularize the inversion of the broadband seismic data, allowing us to recast the interpolation problem as the minimization of the cost function
\begin{equation}
    J(\te) = \Vert \M f_{\te}(\z) - \y \Vert_2^2 + \eps \ani^2 \Vert \Lapl f_{\te}(\z) \Vert_2^2,
    \label{eq:anti-alias}
\end{equation}
where $\ani$ is the vector that collects the anisotropy of the estimated gradient square tensor for each data sample \cite{vanvliet1995estimators}.
In the context of seismic data, $\ani$ can be interpreted as a confidence measure of the estimated slopes. Thus, it is added to the cost function to locally weight the directional Laplacian $\Lapl$.
Notice that, as the optimization goes on, the interpolated data can be used to produce a better estimate of the slopes, updating the Laplacian directions at runtime.

After the minimization of Eq.~\eqref{eq:anti-alias} has been performed, the interpolated solution is obtained as
\begin{equation}
    \x^*=f_{\te^*}\left(\z\right).
\end{equation}

\begin{figure}[t]
    \centering
    \subfloat[\label{fig:lines_input}]{\includegraphics[width=0.25\columnwidth]{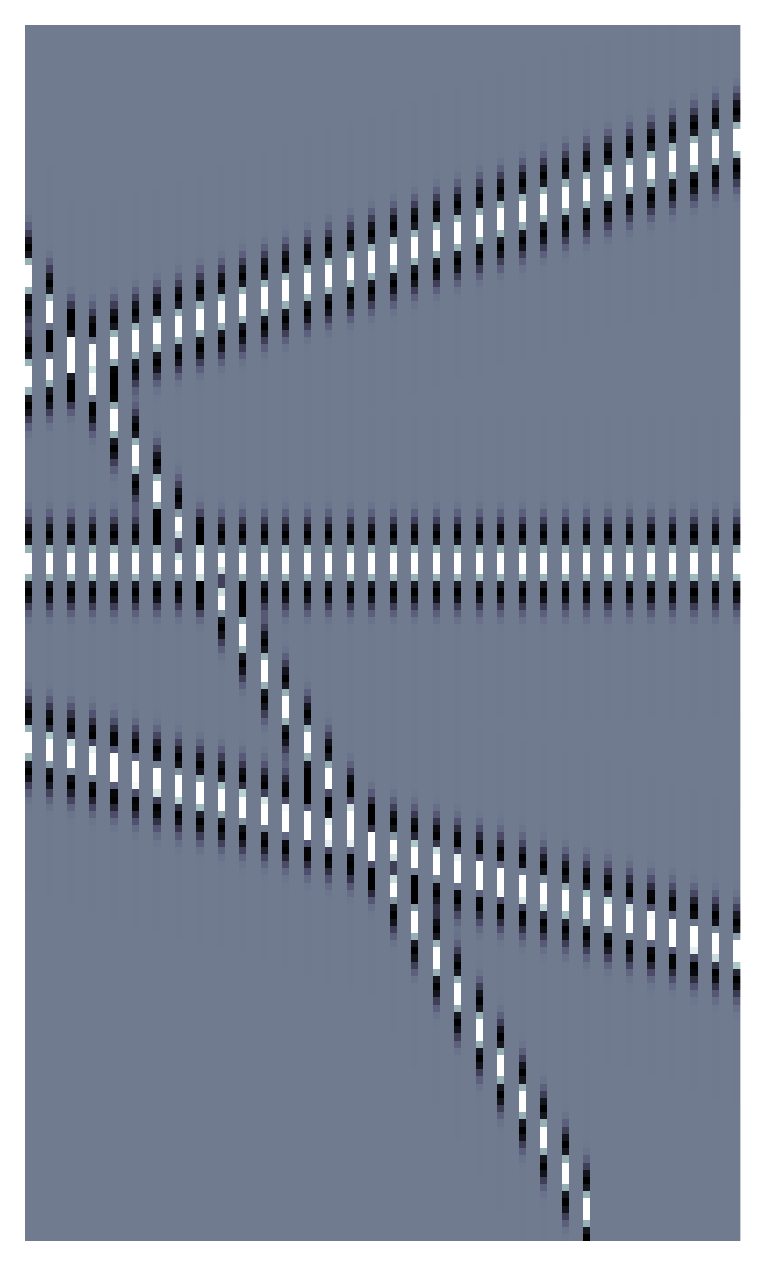}}
    %\subfloat[\label{fig:lines_input}]{\includegraphics[width=0.25\columnwidth]{figure/lines_decimated.pdf}}
    \subfloat[\label{fig:lines_target}]{\includegraphics[width=0.25\columnwidth]{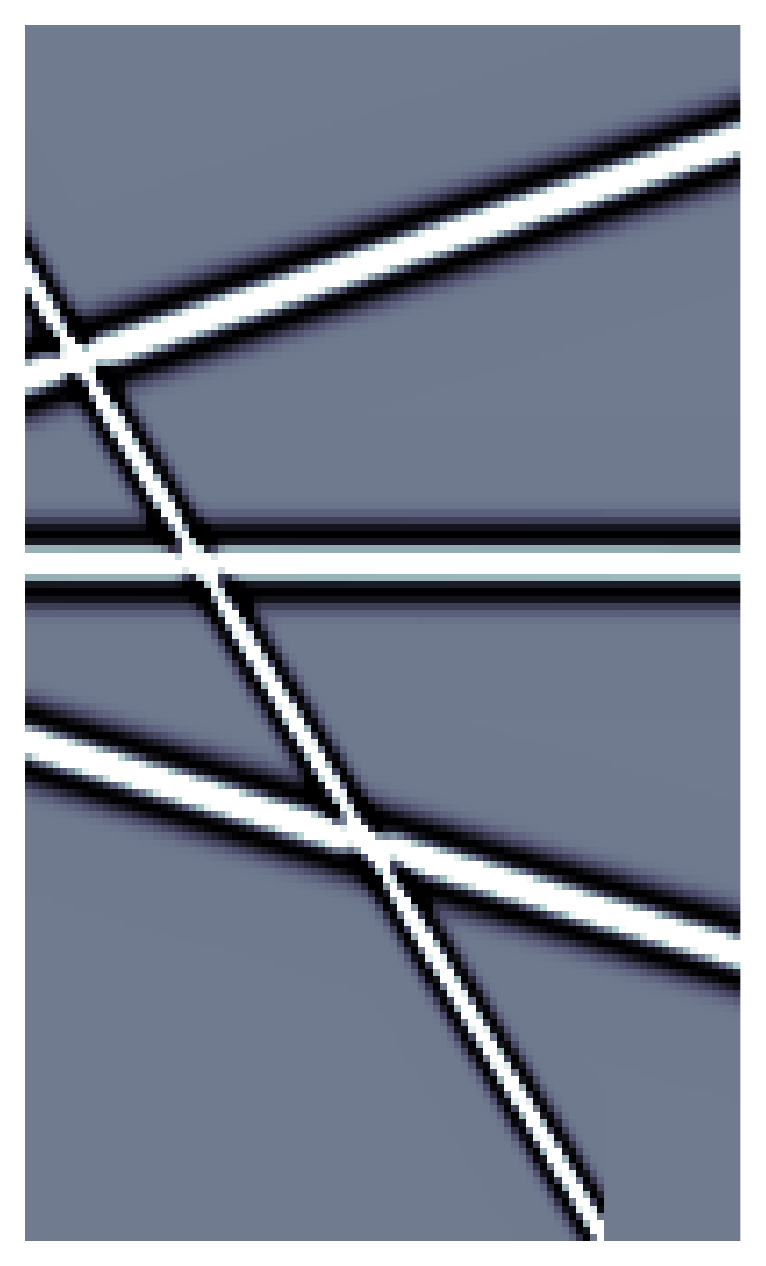}}
    \subfloat[\label{fig:lines_out}]{\includegraphics[width=0.25\columnwidth]{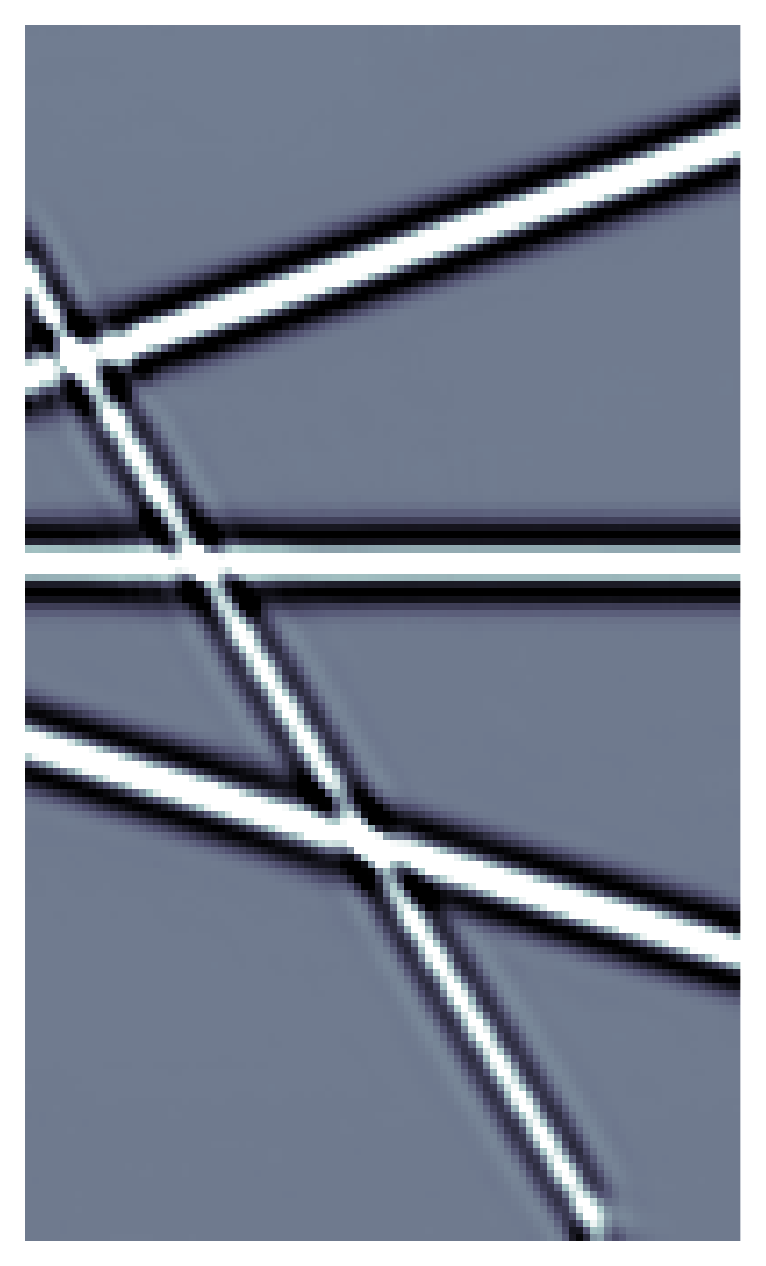}}
    \subfloat[\label{fig:lines_out_dip}]{\includegraphics[width=0.25\columnwidth]{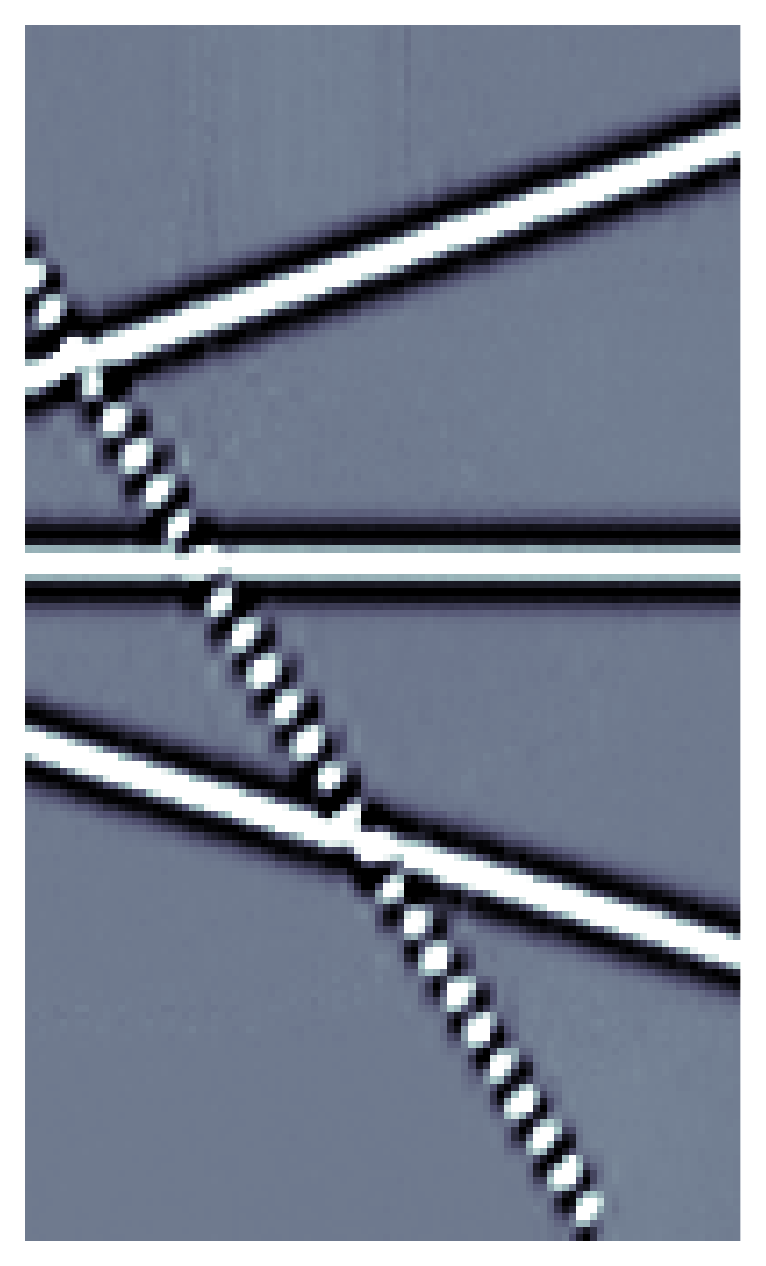}}
    %\subfloat[]{\includegraphics[width=0.15\columnwidth]{figure/lines_error.pdf}}
    %\subfloat[]{\includegraphics[width=0.15\columnwidth]{figure/lines_error_trad.pdf}}
    
    \caption{Results on synthetic linear events: input decimated data (a), reference full data (b), results of the proposed method (c) and standard Deep Prior optimization (d).}
    \label{fig:lines}
\end{figure}

\begin{figure}[t]
    %\vspace{-0.5cm}
    \centering
    %\subfloat[\label{fig:lines_fk_input}]{\includegraphics[width=0.25\columnwidth]{figure/lines_fk_input.pdf}}
    \subfloat[\label{fig:lines_fk_input}]{\includegraphics[width=0.25\columnwidth]{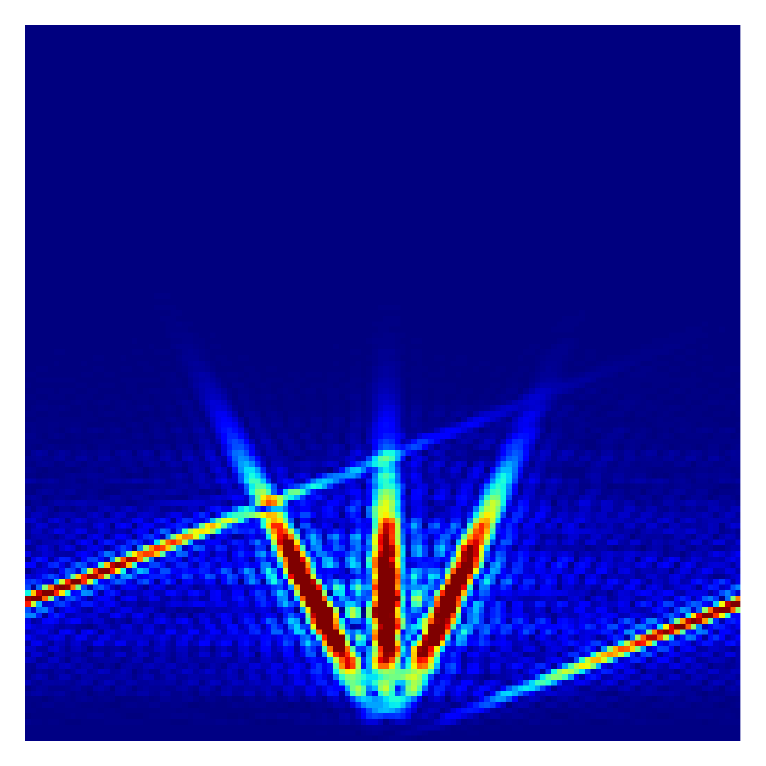}}
    \subfloat[\label{fig:lines_fk_target}]{\includegraphics[width=0.25\columnwidth]{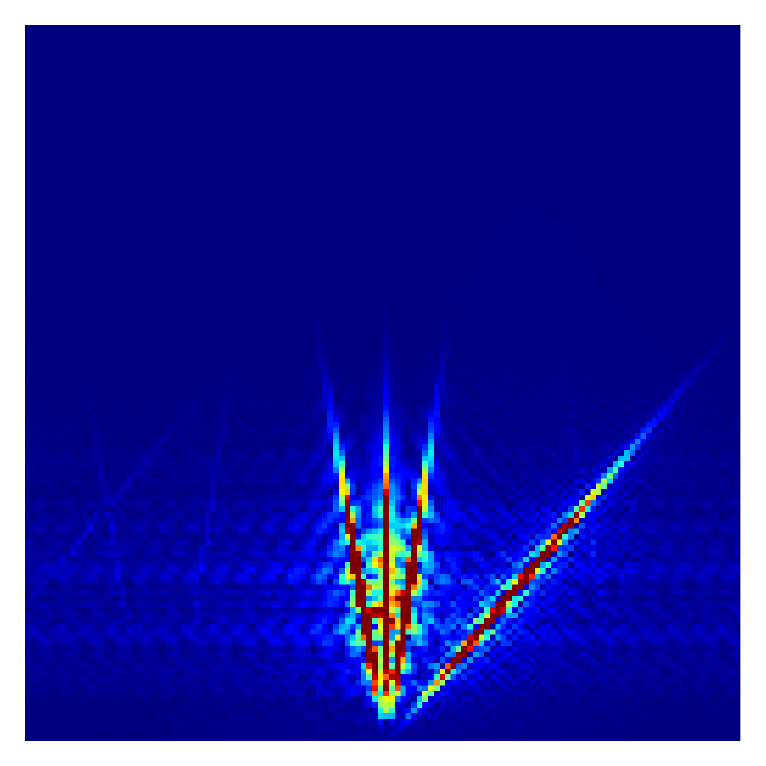}}
    \subfloat[\label{fig:lines_fk_out}]{\includegraphics[width=0.25\columnwidth]{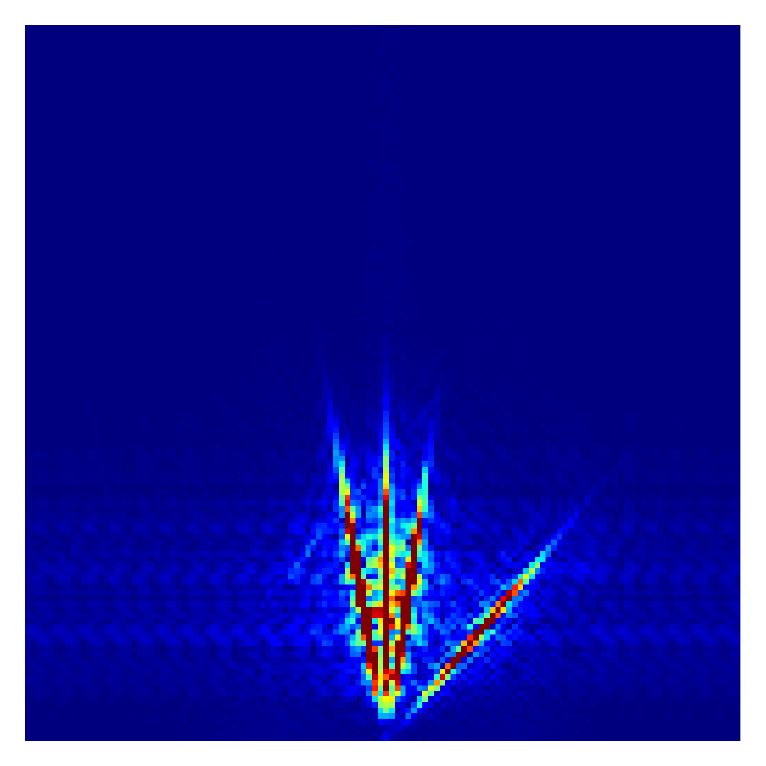}}
    \subfloat[\label{fig:lines_fk_out_dip}]{\includegraphics[width=0.25\columnwidth]{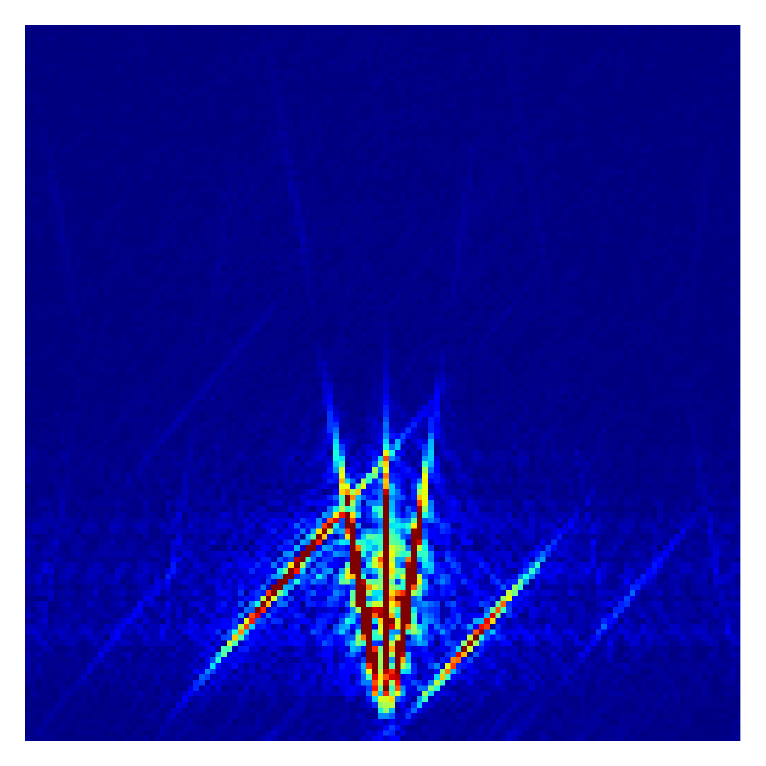}}
    %\subfloat[]{\includegraphics[width=0.15\textwidth]{figure/lines_fk_error.pdf}}
    %\subfloat[]{\includegraphics[width=0.15\textwidth]{figure/lines_fk_error_trad.pdf}}
    
    \caption{FK spectra of synthetic linear events example: input decimated data (a), reference full data (b), results of the proposed method (c) and standard Deep Prior optimization (d).}
    \label{fig:lines_fk}
\end{figure}

\begin{figure*}
    \centering
    \subfloat[\label{fig:f3_input}]{\includegraphics[width=0.16\textwidth]{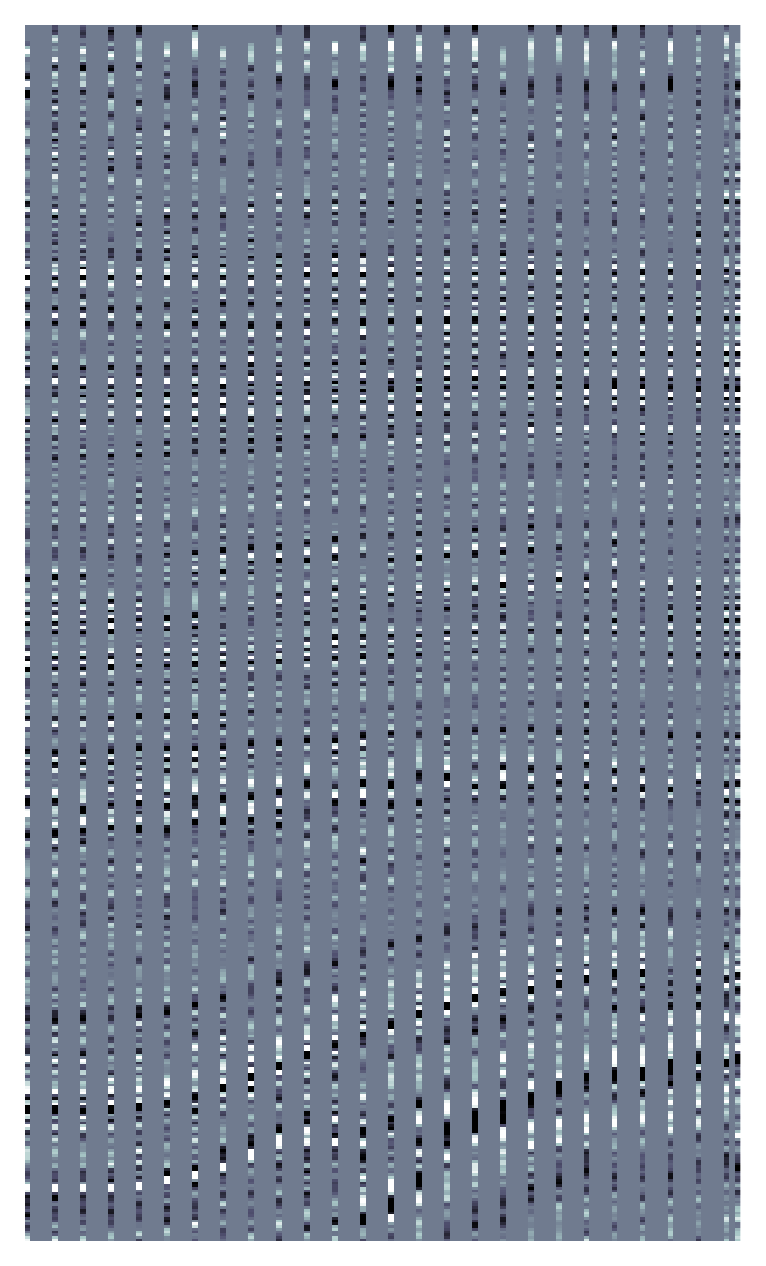}}
    %\subfloat[\label{fig:f3_input}]{\includegraphics[width=0.16\textwidth]{figure/f3_decimated.pdf}}
    \subfloat[\label{fig:f3_target}]{\includegraphics[width=0.16\textwidth]{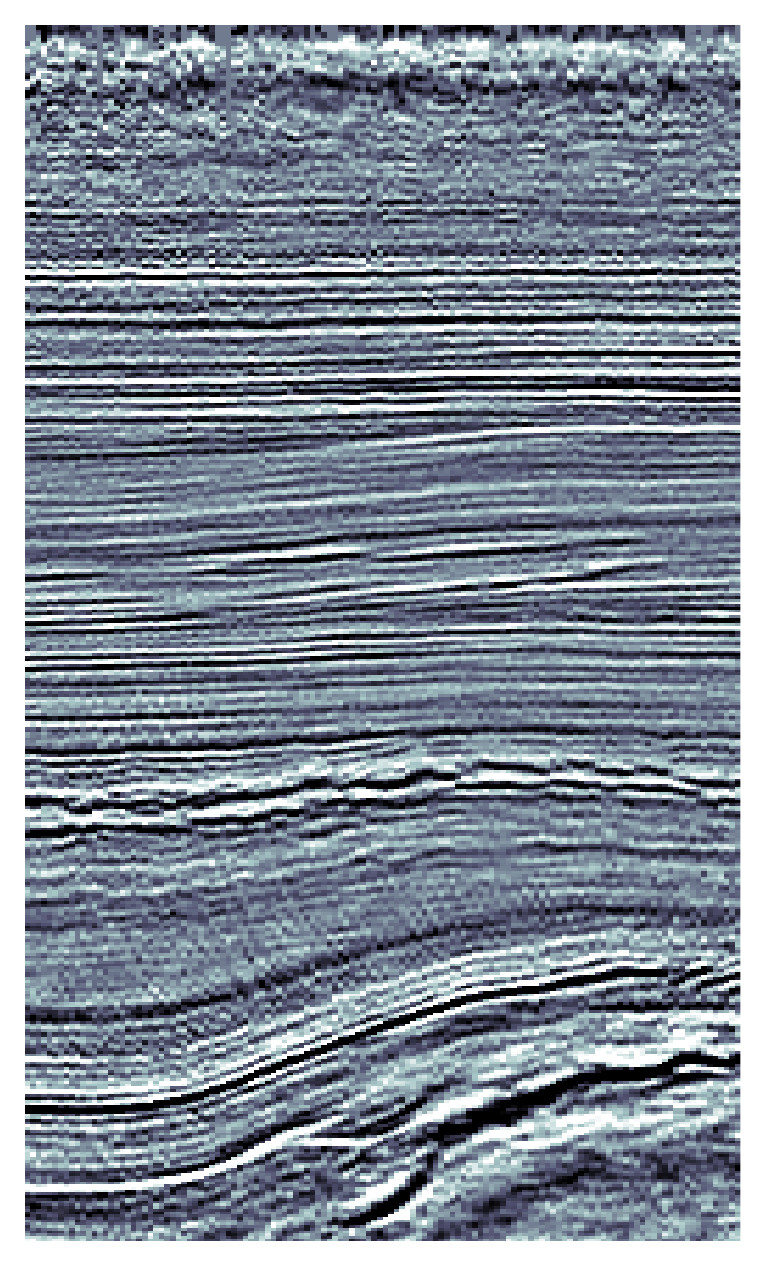}}
    \subfloat[\label{fig:f3_out}]{\includegraphics[width=0.16\textwidth]{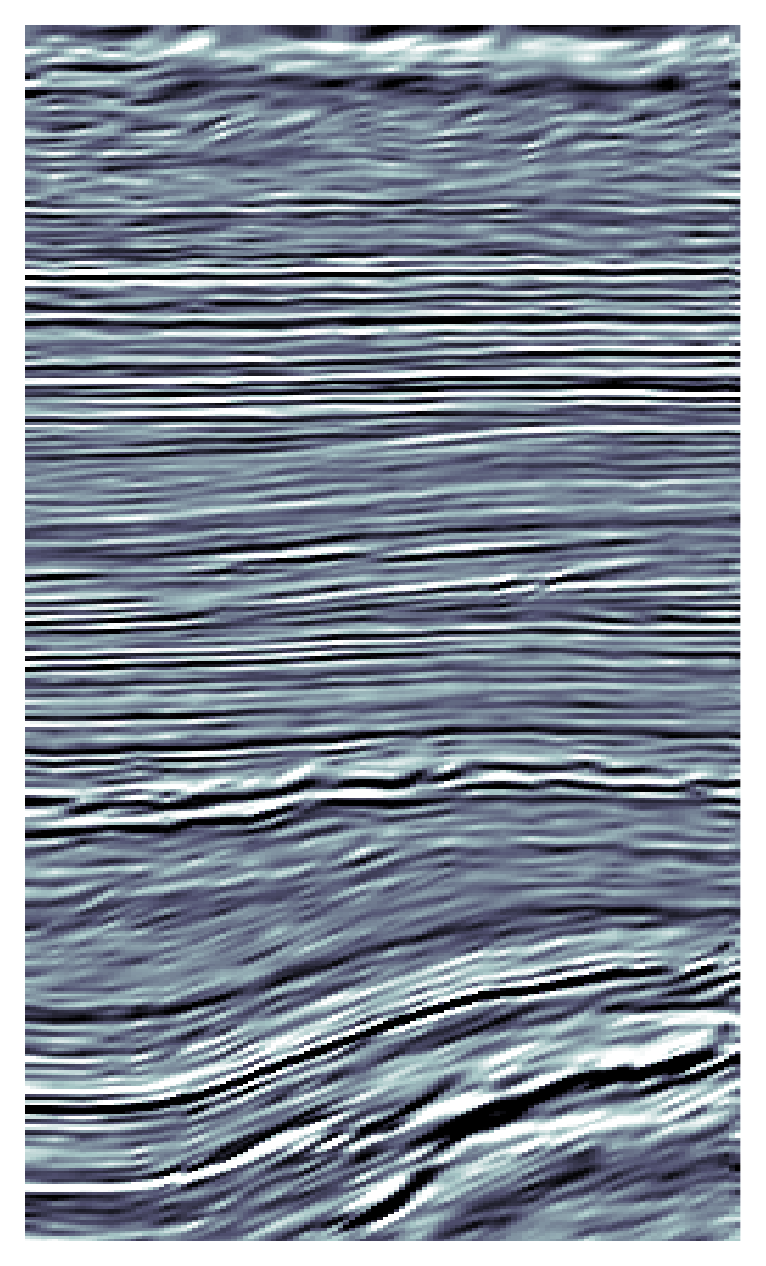}}
    \subfloat[\label{fig:f3_out_dip}]{\includegraphics[width=0.16\textwidth]{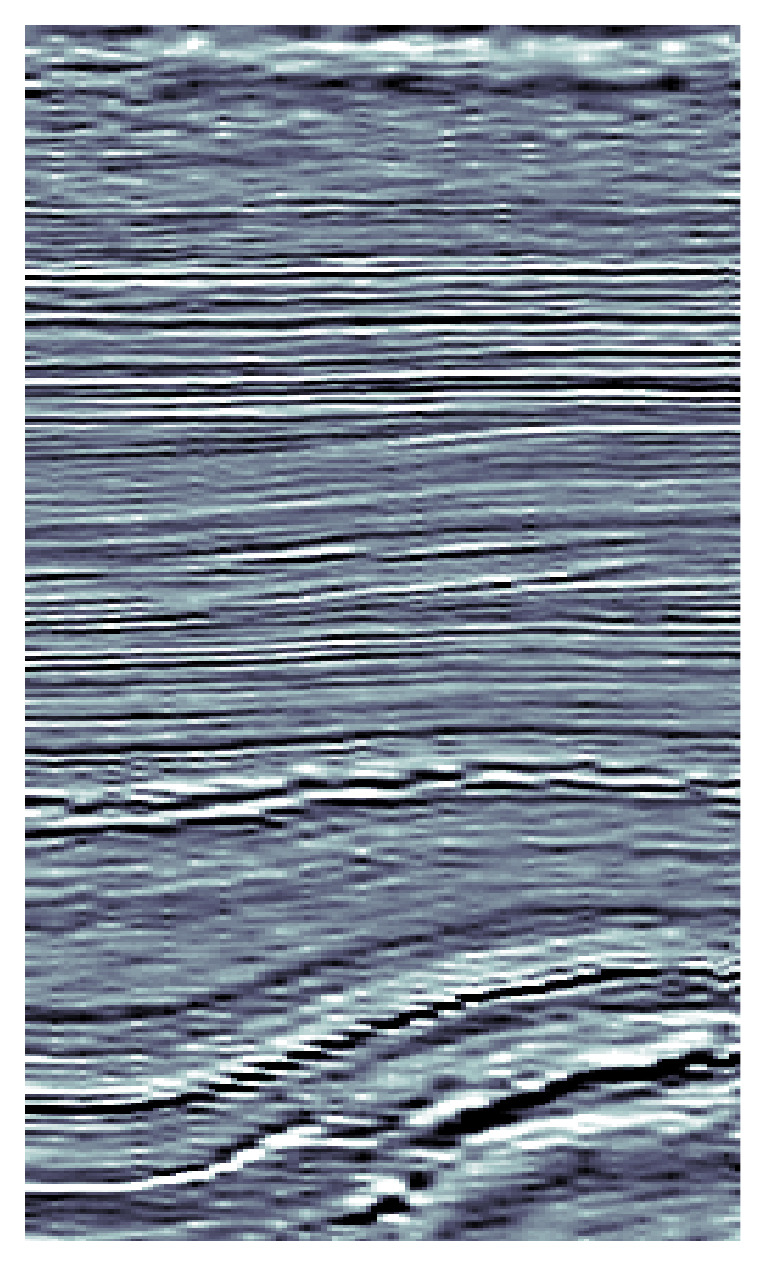}}
    \subfloat[\label{fig:f3_error}]{\includegraphics[width=0.16\textwidth]{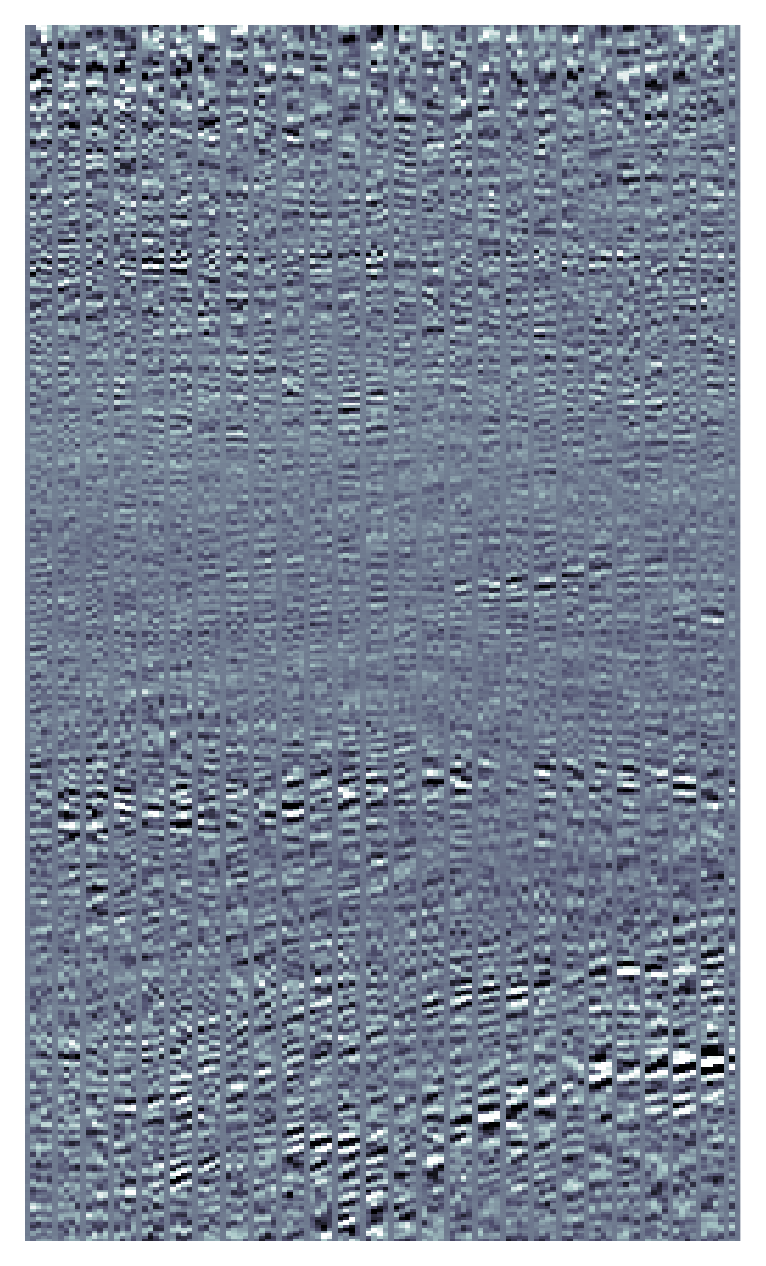}}
    \subfloat[\label{fig:f3_error_dip}]{\includegraphics[width=0.16\textwidth]{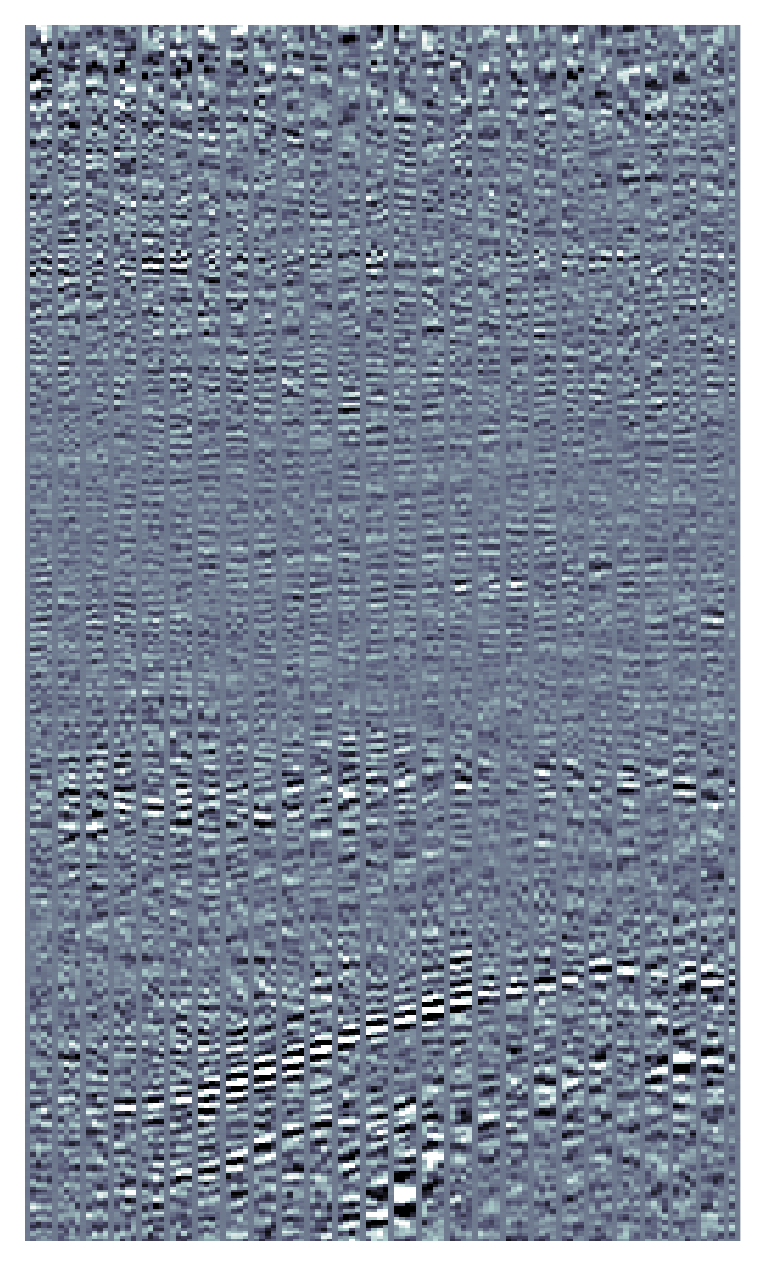}}
    
    \caption{Results on post-stack field data: input decimated data (a), reference full data (b), results of the proposed method (c) and standard Deep Prior optimization (d), along with the reconstruction errors (e, f).}
    \label{fig:f3}
\end{figure*}

\begin{figure}[!t]
    \vspace{-0.5cm}
    \centering
    %\subfloat[\label{fig:f3_fk_input}]{\includegraphics[width=0.25\columnwidth]{figure/f3_fk_input.pdf}}
    \subfloat[\label{fig:f3_fk_input}]{\includegraphics[width=0.25\columnwidth]{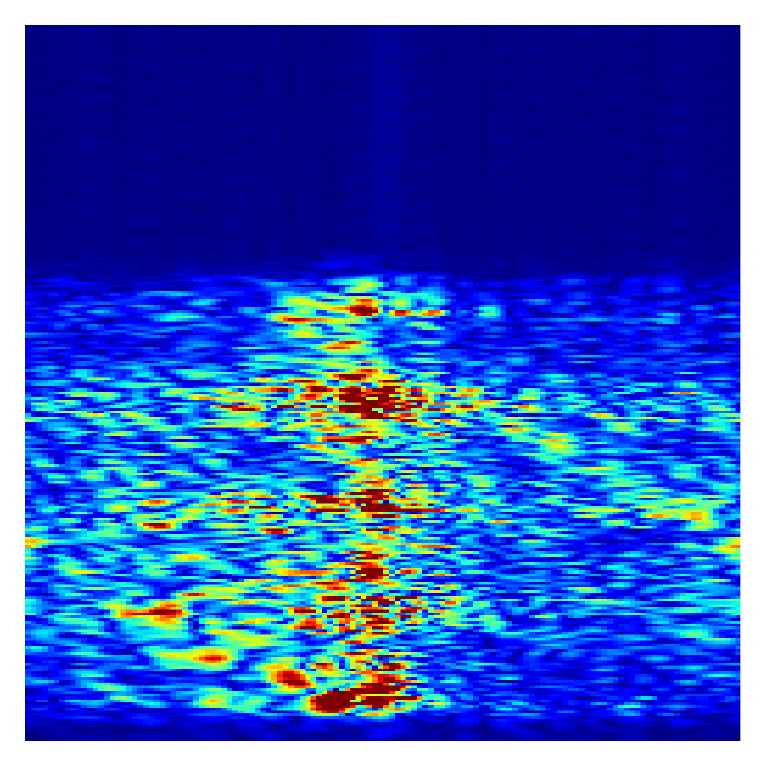}}
    \subfloat[\label{fig:f3_fk_target}]{\includegraphics[width=0.25\columnwidth]{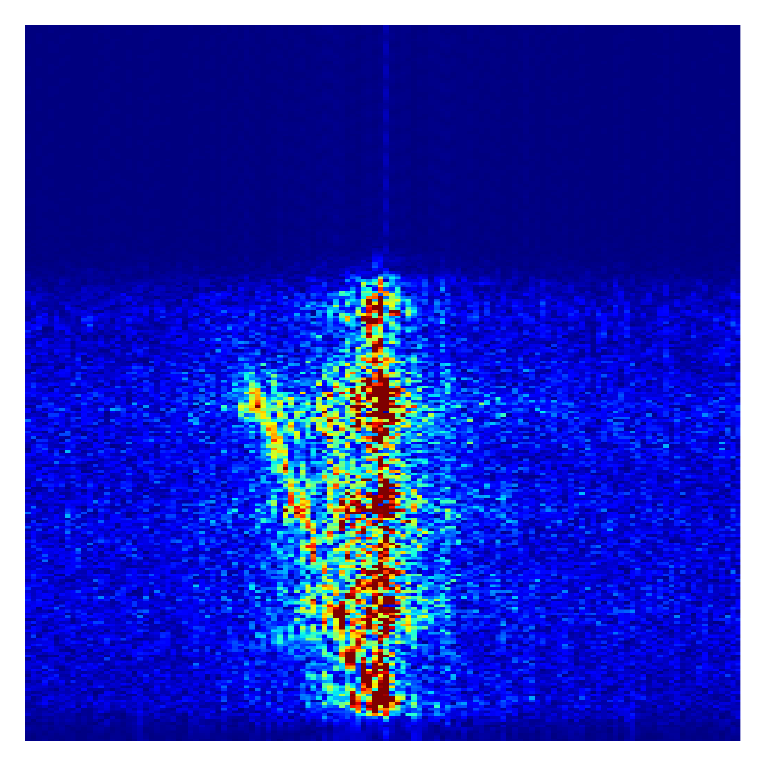}}
    \subfloat[\label{fig:f3_fk_out}]{\includegraphics[width=0.25\columnwidth]{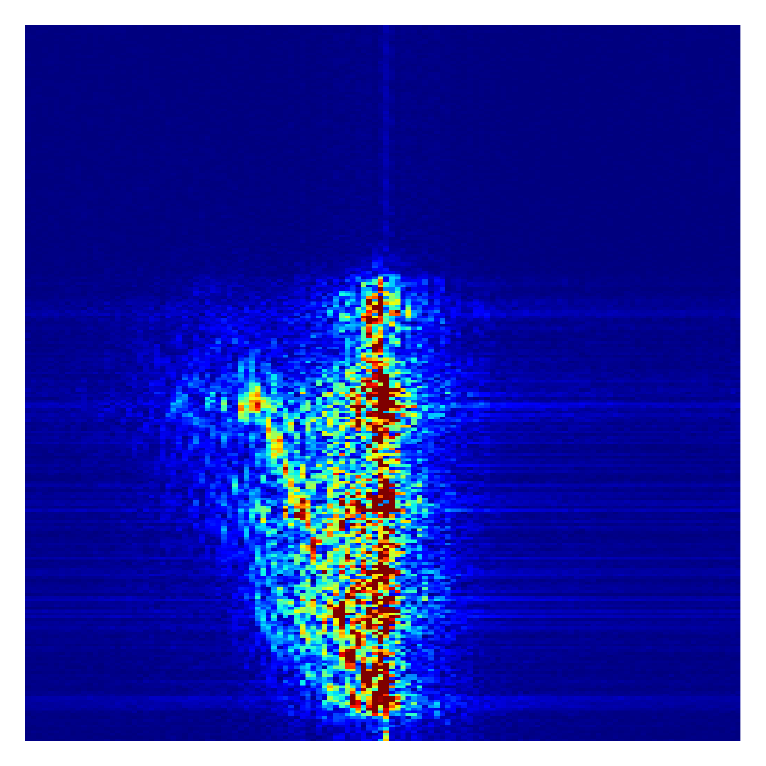}}
    \subfloat[\label{fig:f3_fk_out_dip}]{\includegraphics[width=0.25\columnwidth]{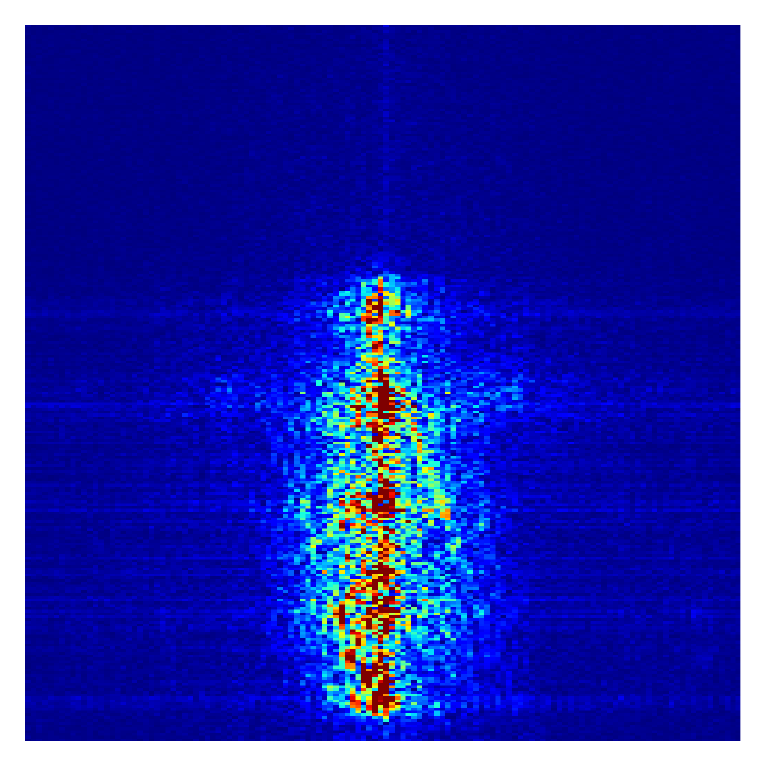}}
    %\subfloat[]{\includegraphics[width=0.15\textwidth]{figure/lines_fk_error.pdf}}
    %\subfloat[]{\includegraphics[width=0.15\textwidth]{figure/lines_fk_error_trad.pdf}}
    
    \caption{FK spectra of post-stack field data: input decimated data (a), reference full data (b), results of the proposed method (c) and standard Deep Prior optimization (d).}
    \label{fig:f3_fk}
\end{figure}
\begin{figure}[!h]
    \vspace{-0.55cm}
    \centering
    \subfloat[\label{fig:f3_dip}]{\includegraphics[width=0.19\textwidth]{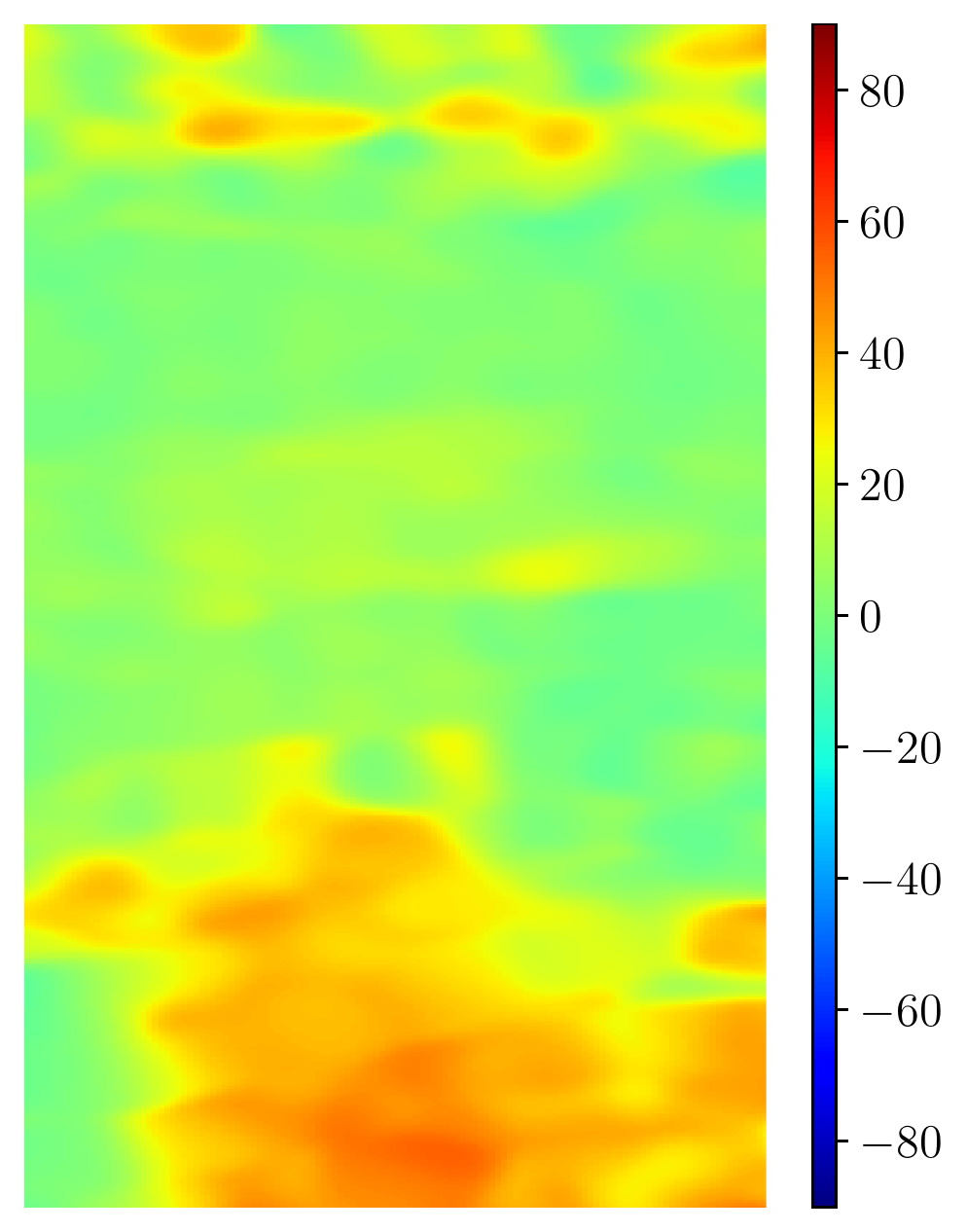}}
    \hspace{1cm}
    \subfloat[\label{fig:f3_ani}]{\includegraphics[width=0.19\textwidth]{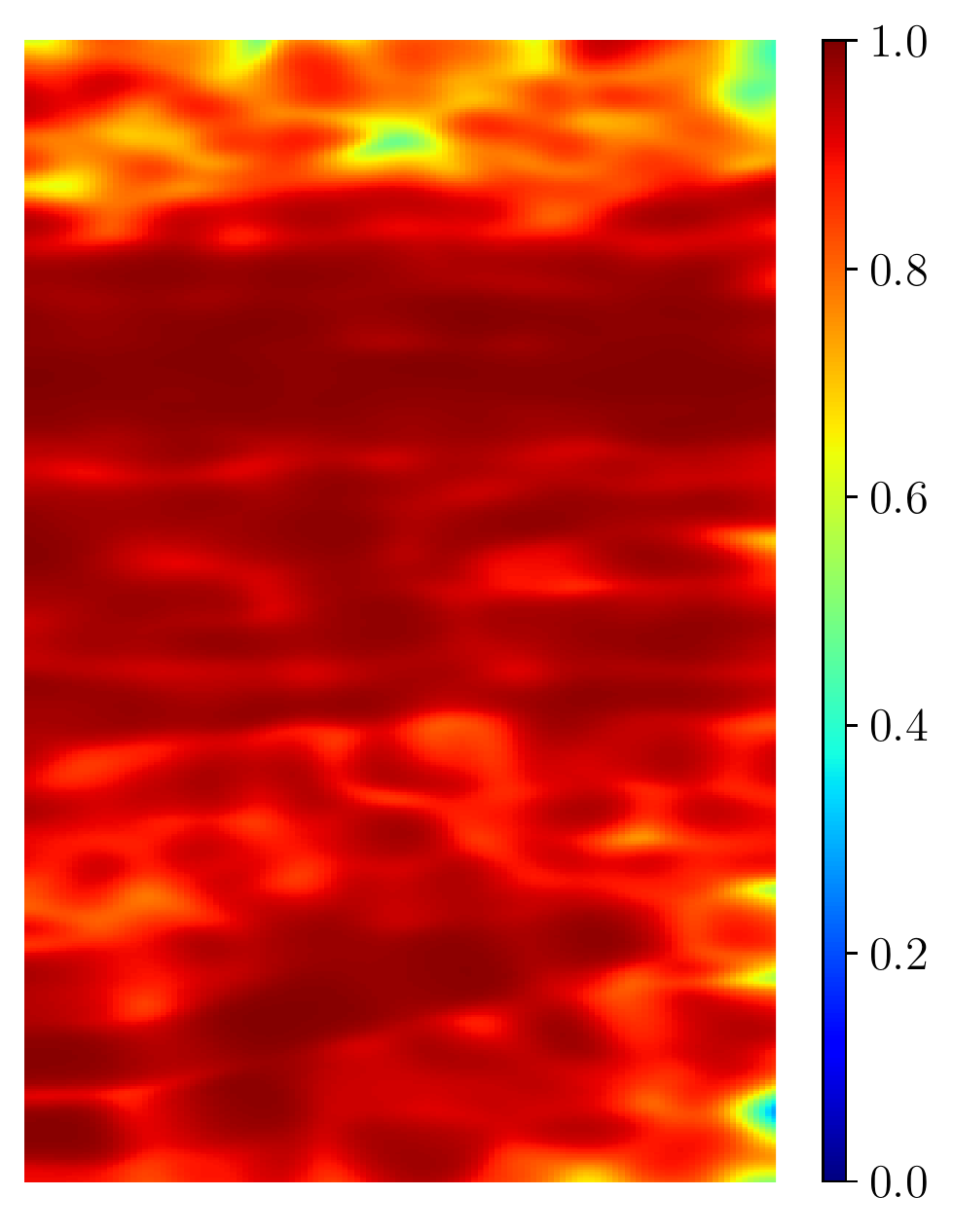}}
    \caption{Slopes $\slopes$ in degree (a) and related confidence $\ani$ (b) estimated on the interpolated F3 data.}
    \label{fig:f3_slopes}
\end{figure}
\begin{figure}[!t]
    \centering
    \begin{tikzpicture}
        %\hspace{-0.5cm}
        \begin{axis}[
            height=4cm,
            width=\columnwidth, 
            xlabel={Iterations},
            xmin=0, xmax=6000,
            ymin=0.00001, ymax=0.01,
            ymode=log,
            legend pos=north east,
            ymajorgrids=true,
            xmajorgrids=true,
            grid style=dashed,
            legend cell align={left},
        ]
    
        \addplot[mark=none, blue] table[x=iteration, y=datafidelity]{f3.txt};
    
        \addplot[mark=none, orange] table[x=iteration, y=laplacian]{f3.txt};
        \legend{$\Vert \M f_{\te}(\z) - \y \Vert_2^2$,
                $\Vert \Lapl f_{\te}(\z) \Vert_2^2$}
        \end{axis}
    \end{tikzpicture}
    \caption{Objective function components of post-stack field data interpolation.}
    \label{fig:f3_curve}
\end{figure}
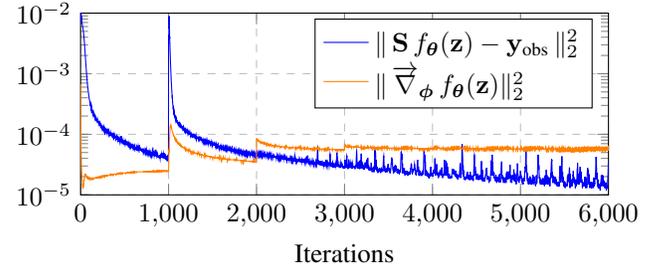

\begin{figure*}[!t]
    \centering
    \subfloat[\label{fig:viking_input}]{\includegraphics[width=0.16\textwidth]{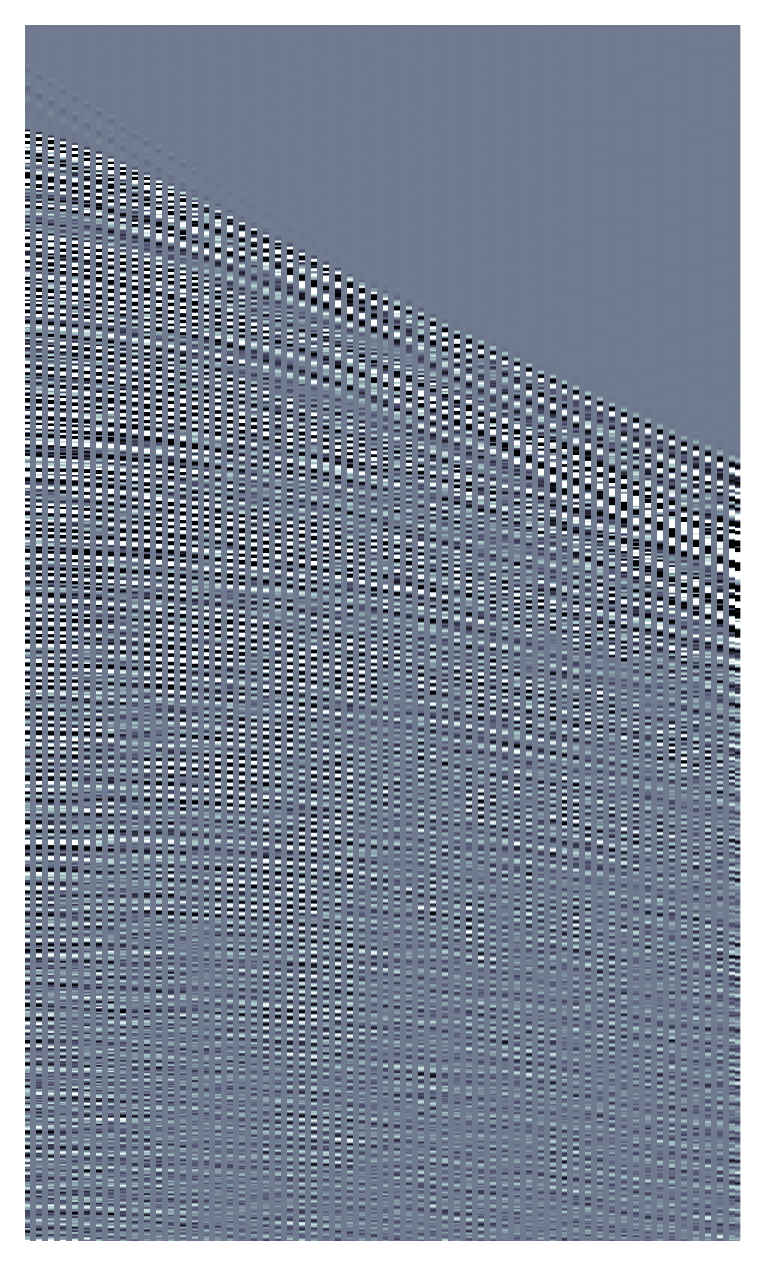}}
    %\subfloat[\label{fig:viking_input}]{\includegraphics[width=0.16\textwidth]{figure/viking_decimated.pdf}}
    \subfloat[\label{fig:viking_target}]{\includegraphics[width=0.16\textwidth]{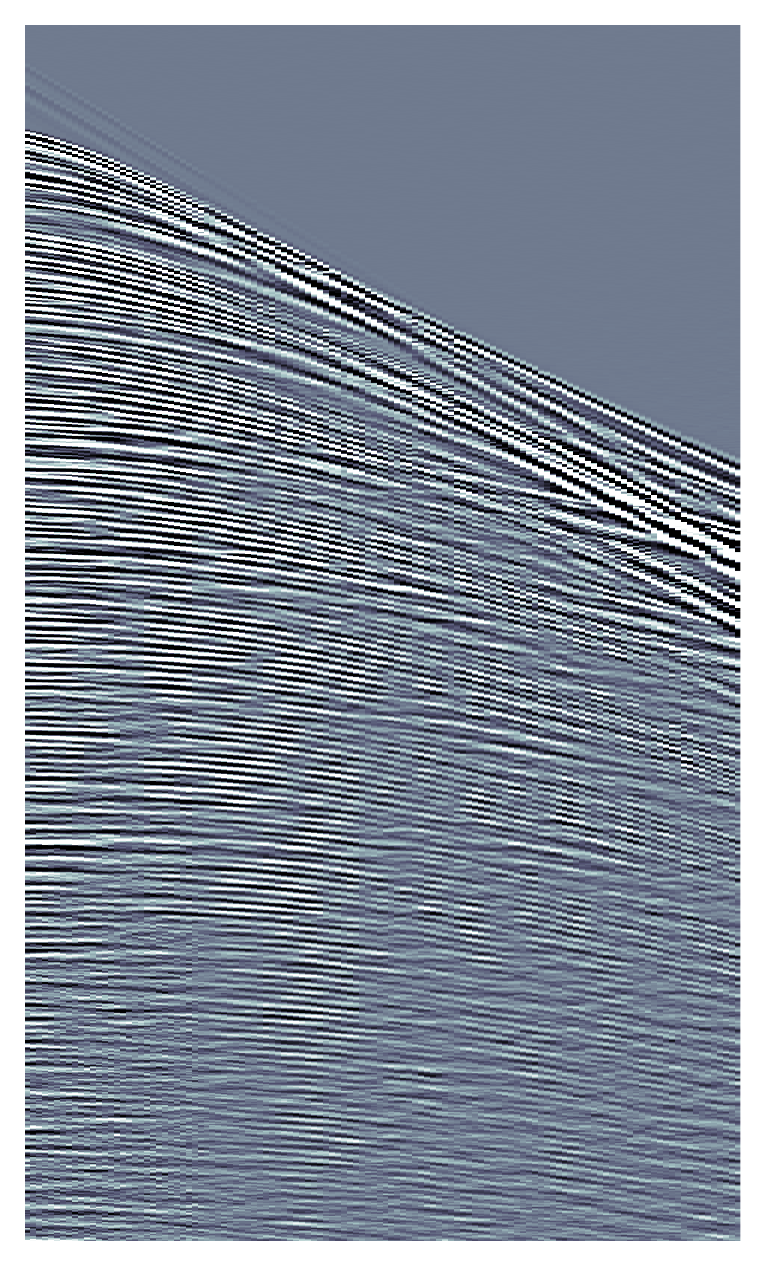}}
    \subfloat[\label{fig:viking_out}]{\includegraphics[width=0.16\textwidth]{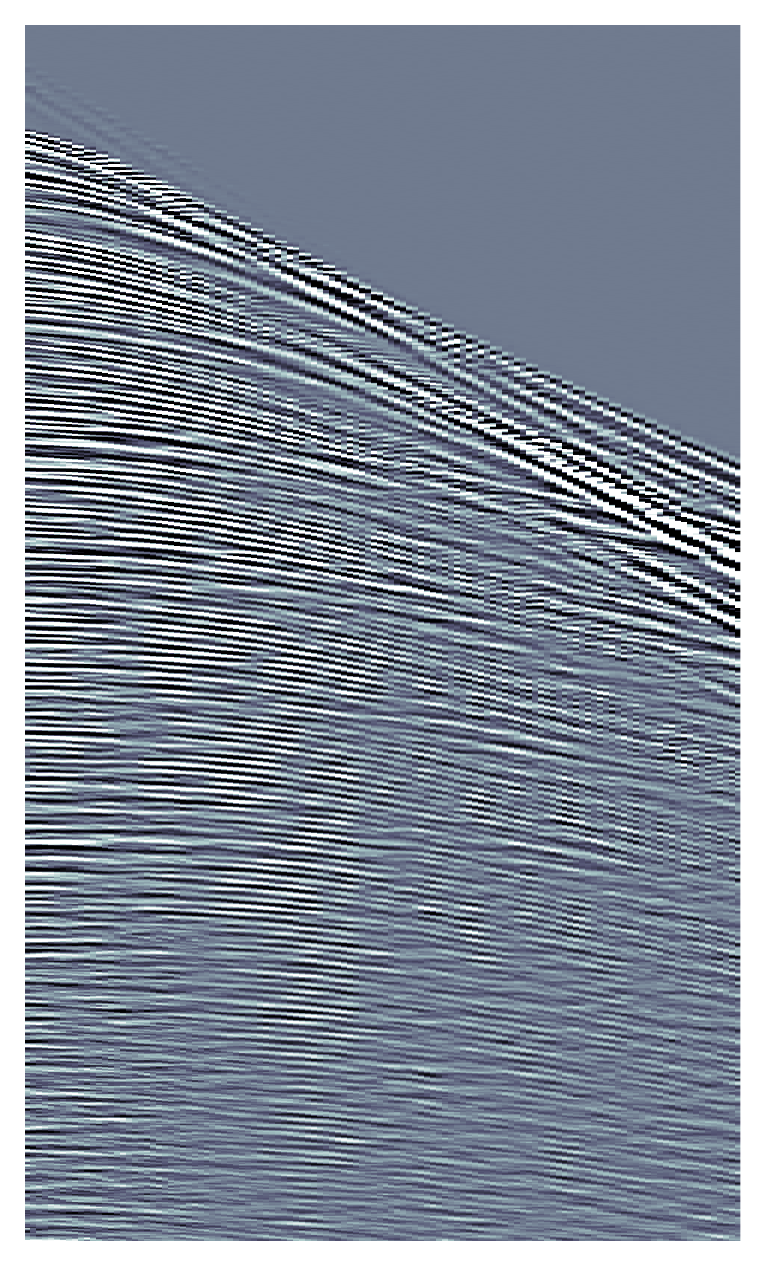}}
    \subfloat[\label{fig:viking_out_dip}]{\includegraphics[width=0.16\textwidth]{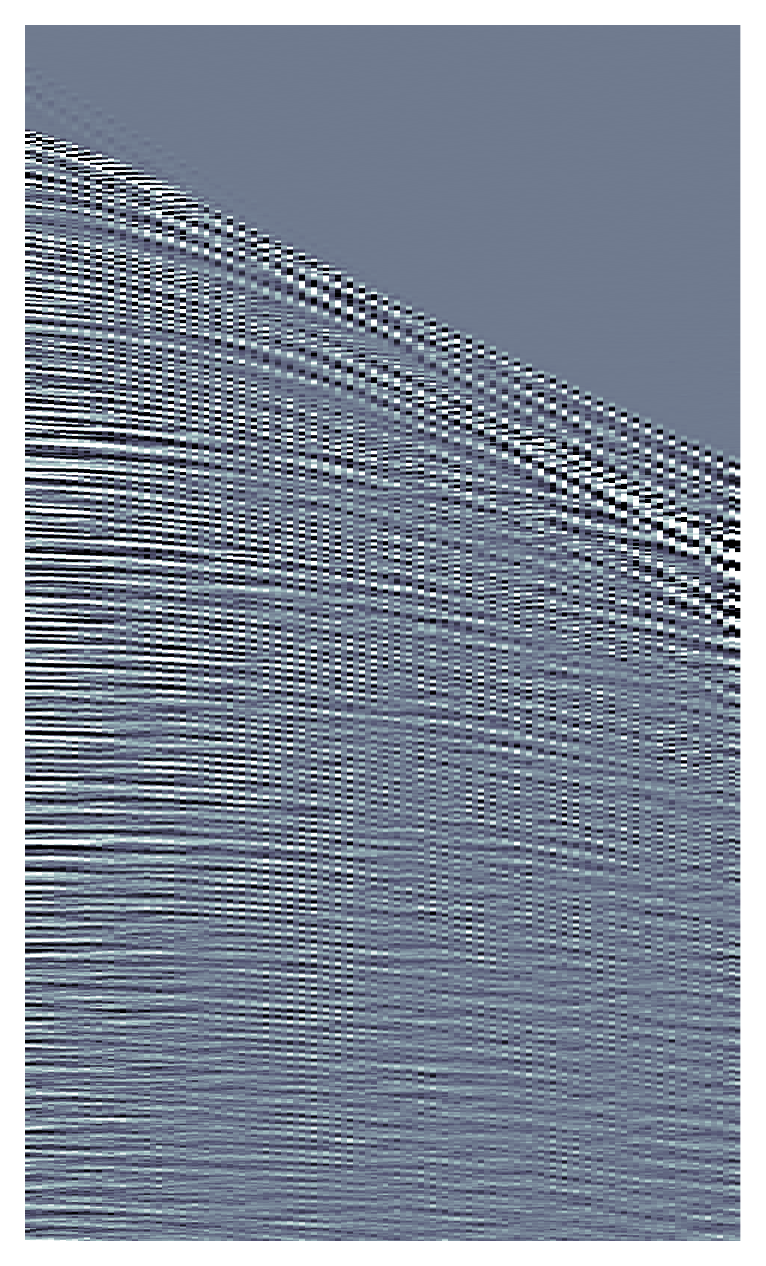}}
    \subfloat[\label{fig:viking_error}]{\includegraphics[width=0.16\textwidth]{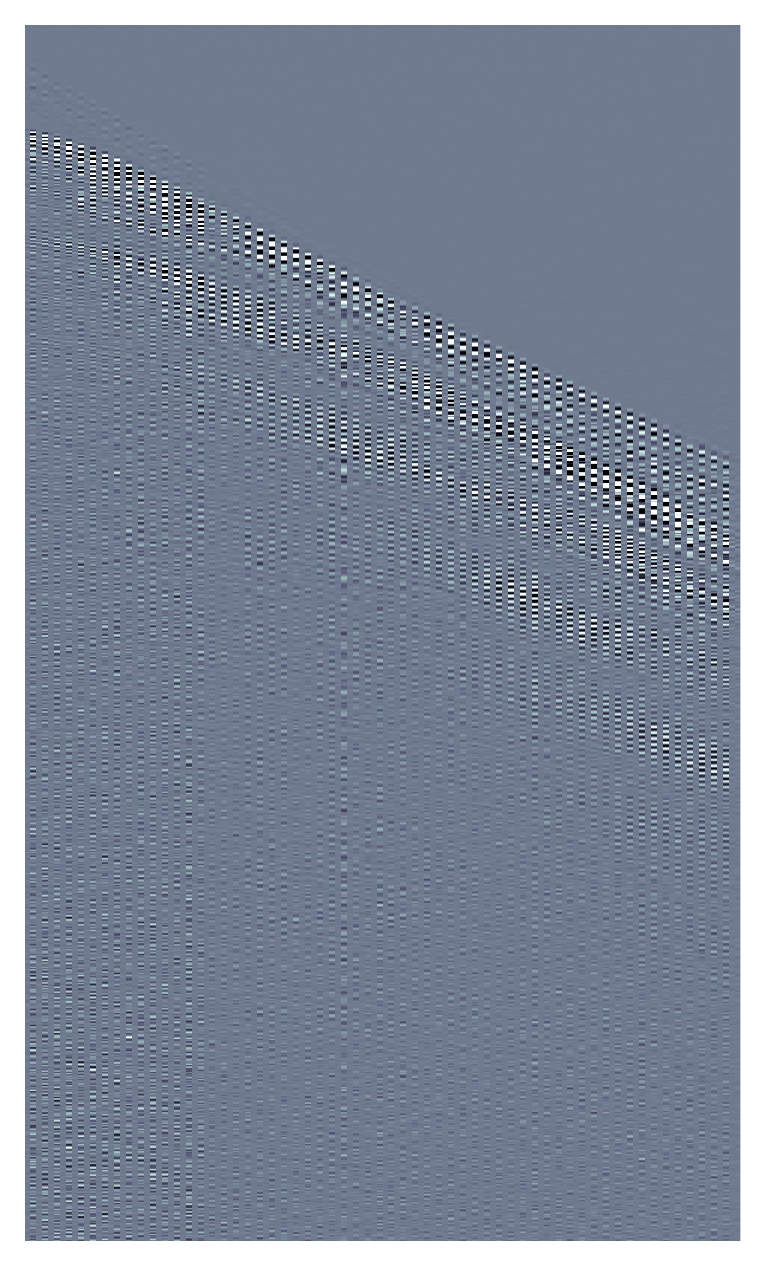}}
    \subfloat[\label{fig:viking_error_dip}]{\includegraphics[width=0.16\textwidth]{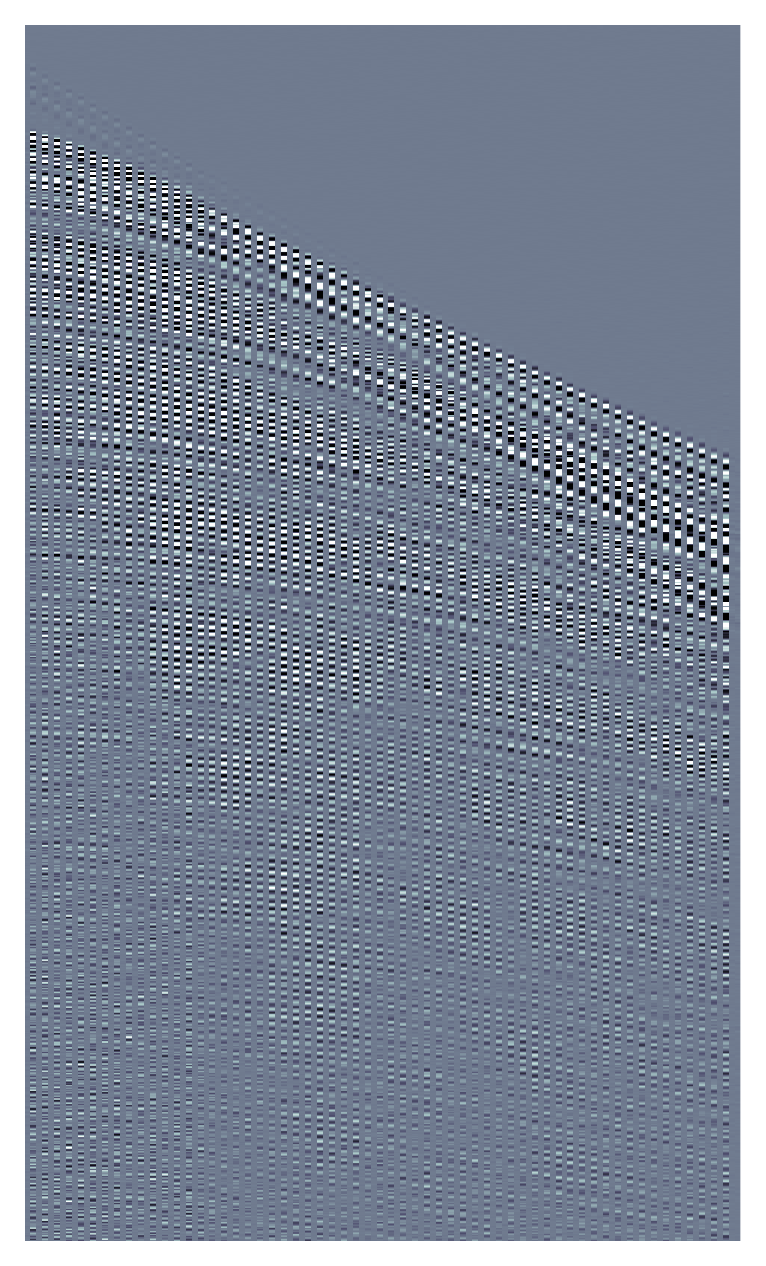}}
    
    \caption{Results on pre-stack field data: input decimated data (a), reference full data (b), results of the proposed method (c) and standard Deep Prior optimization (d), along with the reconstruction errors (e, f).}
    \label{fig:viking}
\end{figure*}

\begin{figure}[!t]
    \vspace{-0.3cm}
    \centering
    %\subfloat[\label{fig:viking_fk_input}]{\includegraphics[width=0.25\columnwidth]{figure/viking_fk_input.pdf}}
    \subfloat[\label{fig:viking_fk_input}]{\includegraphics[width=0.25\columnwidth]{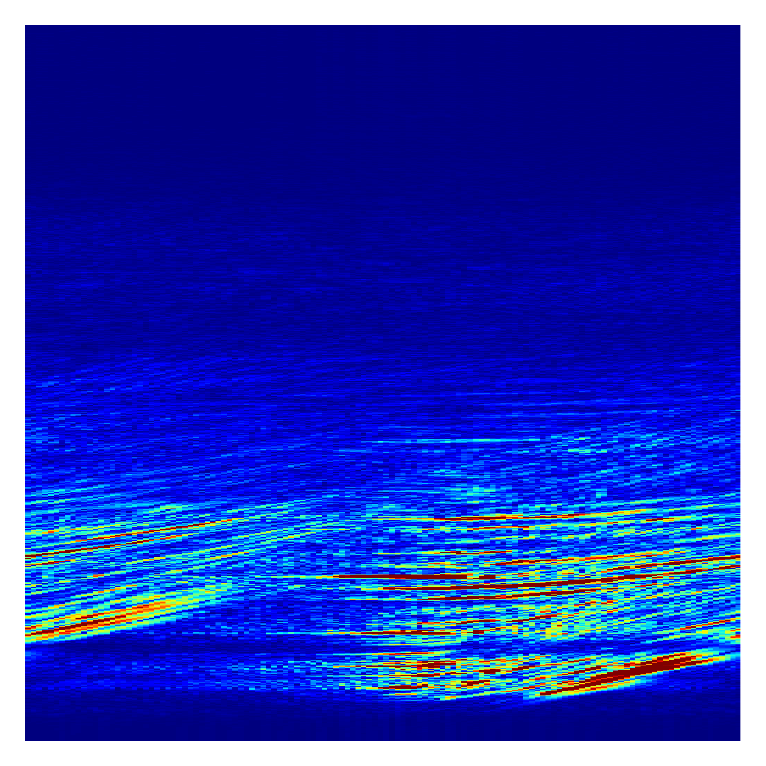}}
    \subfloat[\label{fig:viking_fk_target}]{\includegraphics[width=0.25\columnwidth]{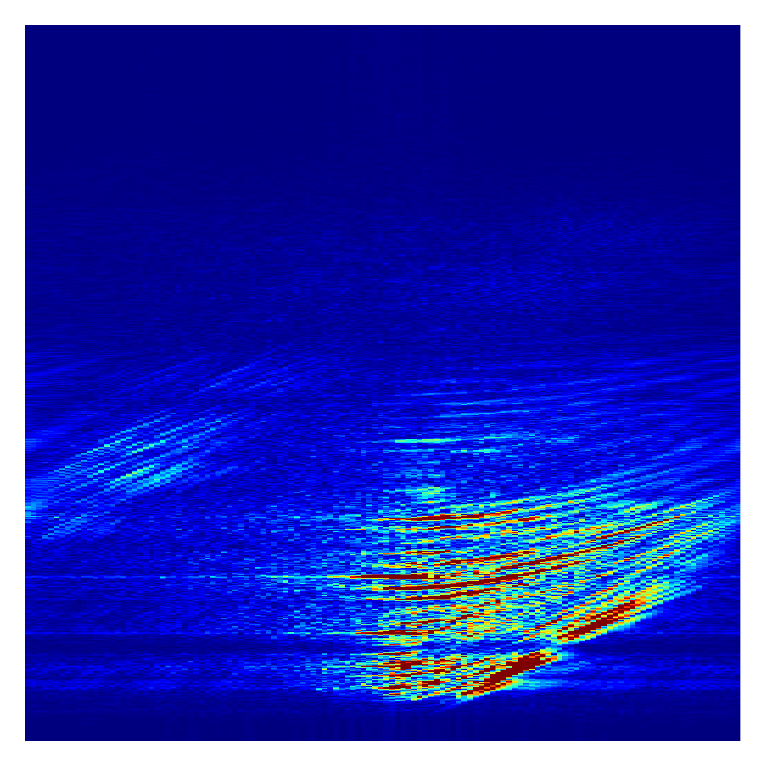}}
    \subfloat[\label{fig:viking_fk_out}]{\includegraphics[width=0.25\columnwidth]{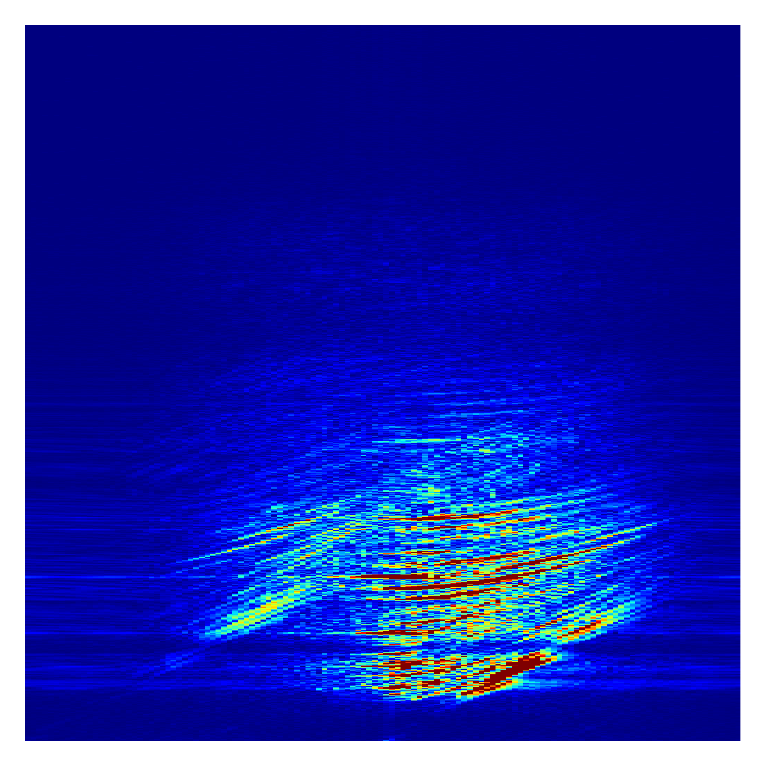}}
    \subfloat[\label{fig:viking_fk_out_dip}]{\includegraphics[width=0.25\columnwidth]{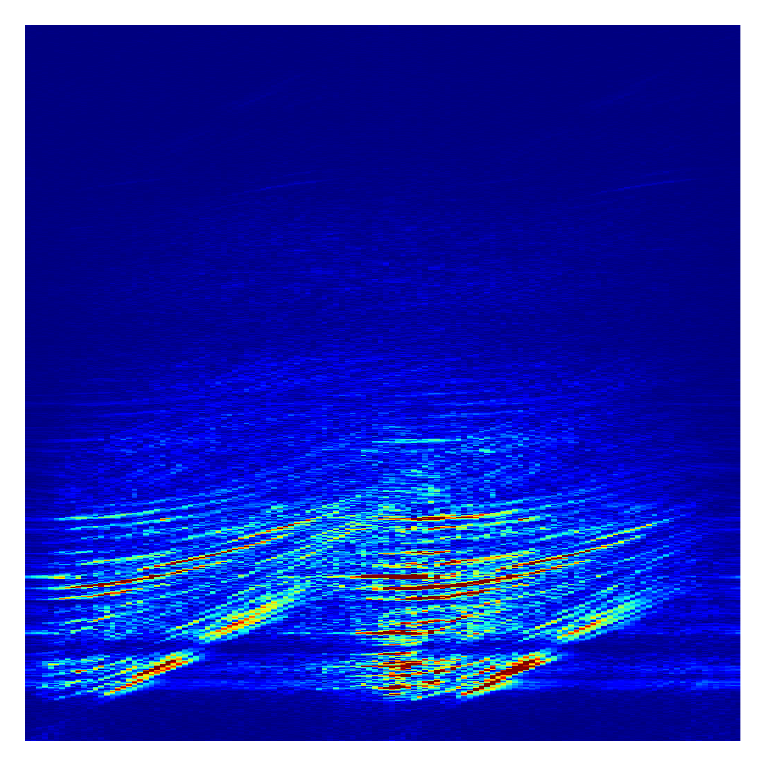}}
    %\subfloat[]{\includegraphics[width=0.15\textwidth]{figure/lines_fk_error.pdf}}
    %\subfloat[]{\includegraphics[width=0.15\textwidth]{figure/lines_fk_error_trad.pdf}}
    
    \caption{FK spectra of pre-stack field data: input decimated data (a), reference full data (b), results of the proposed method (c) and standard Deep Prior optimization (d).}
    \label{fig:viking_fk}
\end{figure}

\section{Numerical Results}\label{sec:results}
In this section, we show numerical results obtained with our methodology on three different use cases. The CNN architecture is a standard U-Net design with skip connections. The input noise $\z$ is sampled from a Gaussian distribution $\mathcal{N}(0, 0.1)$. Optimization is performed through Adam algorithm with learning rate $0.001$. Low-pass filtering is performed through a second order Butterworth design.
As a reconstruction metrics, we compute the total SNR of the interpolated data with respect to the reference data. We underline the fact that the latter is not included in the optimization process: the network is not fed with the reference full data at any step.

\subsection{First use-case: synthetic}
First, we employ a synthetic examples made of 4 linear events with different dips. In particular, three events are mildly steep, while the fourth event has been designed to create aliasing effects.
This example is simple yet particularly challenging due to the slope of the events.
The data is made of $170$ time samples with sampling frequency $1$~kHz and $100$ traces equispaced by $5$~m. Data are depicted in Fig.~\ref{fig:lines_target}. 
We build a regular binary mask for simulating the decimation by keeping 1 trace over 3, as depicted in Figure~\ref{fig:lines_input}.
The cutoff frequency $f_c$ is $50$~Hz; the network is optimized for the first $3000$ iterations to fit the low-passed data, as in Eq.~\eqref{eq:DIP_low}. Then, the slopes are estimated and the optimization is performed on the sub-sampled data for $7000$ iterations more with regularization weight $\eps=5$.

Figures~\ref{fig:lines_out} and~\ref{fig:lines_out_dip} show the results of Deep Prior optimization when minimizing the proposed objective \eqref{eq:anti-alias} and the standard data fitting, respectively.
The achieved SNR is $12.59$~dB and $6.99$~dB, respectively.
We can notice that the standard Deep Prior optimization could not resolve the aliasing of the steeper event, while the proposed methodology shows a more homogeneous reconstruction.
This effect is particularly visible also in the spectra depicted in Figure~\ref{fig:lines_fk_out_dip}, where we can notice a residual aliasing pattern; on the other hand, in Figure~\ref{fig:lines_fk_out} it has been removed.

\subsection{Second use-case: post-stack}
Encouraged by this result, we processed a post-stack 2D section of the F3 Netherlands field data \cite{silva2019netherlands}.
This section consists of $448$ time samples, with sampling period $16$~ms, and $128$ traces with $25$~m spacing, as depicted in Figure~\ref{fig:f3_target}.
This example is quite complex, with different slopes and noise residuals coming from previous data processing.
The regular binary mask keeps 1 trace over 5, as depicted in Figure~\ref{fig:f3_input}.
The cutoff frequency $f_c$ is $10$~Hz; the network is optimized for the first $1000$ iterations to fit the low-passed data, as in Eq.~\eqref{eq:DIP_low}. Then, the slopes are estimated and the optimization is performed on the sub-sampled data for $5000$ iterations more with regularization weight $\eps=0.05$.
The behavior of the cost function terms at each iteration can be observed in Figure~\ref{fig:f3_curve}. The big discontinuity in the data fidelity term (in blue) arises because the optimization switches from Eq.~\eqref{eq:DIP_low} to Eq.~\eqref{eq:anti-alias}. The orange curve depicts the Laplacian regularizer. For the first 1000 iterations it has not been taken into account; then, it is included in the optimization. The discontinuities visible at iteration 2000, 3000 etc. are due to the fact that the slopes $\slopes$ and confidence $\ani$ are refined every 1000 iterations from the low-pass interpolated data.

Figures~\ref{fig:f3_out} and~\ref{fig:f3_out_dip} show the results of our approach and a standard Deep Prior optimization, respectively.
Both techniques produce a good estimate of the flat zones.
On the other hand, where the interpolated data in Figure~\ref{fig:f3_out_dip} is aliased, the directional Laplacian regularizer has improved the continuity of the steeper reflections.
The total SNR of the proposed approach is $6.62$~dB, against $5.57$~dB obtained without regularization.
Moreover, the Deep Prior paradigm produces estimates of the data with less noise content; this is clearly visible in the spectrum depicted in Figure~\ref{fig:f3_fk_out}.

Figure~\ref{fig:f3_slopes} depicts the slopes $\slopes$ and its confidence $\ani$ estimated from the low-pass filtered interpolated data. By smoothing the gradient with $\sigma=5$, we are able to discriminate local dips with a good resolution.

\subsection{Third use-case: pre-stack}
Finally, we show the limitations of the devised methodology on pre-stack seismic data taken from the Viking Graeben Line 12. This dataset consists of $1408$ time samples, with sampling period $4$~ms, and $120$ traces with $25$~m spacing. Moreover, the dataset is affected by noise and aliasing, visible in Figure~\ref{fig:viking_target} and its spectrum in Figure~\ref{fig:viking_fk_target}.
In order to equalize the amplitude of the late events, we apply a linear gain on each trace.
The decimation is $50\%$, as depicted in Figure~\ref{fig:viking_input}. 
The cutoff frequency $f_c$ is $20$~Hz; the network is optimized for the first $2000$ iterations to fit the low-passed data, as in Eq.~\eqref{eq:DIP_low}. Then, the slopes are estimated and the optimization is performed on the sub-sampled data for $6000$ iterations more with regularization weight $\eps=0.05$.

This example shows clearly the limitation of the standard Deep Prior inversion. Almost every event is poorly reconstructed, as depicted in Figure~\ref{fig:viking_out_dip}. The achieved SNR is $4.47$~dB.
Adopting the proposed technique, data in Figure~\ref{fig:viking_out} are better interpolated; the total SNR is $8.34$~dB.
The residual depicted in Figure~\ref{fig:viking_error} is concentrated almost exclusively on the first events.
However, a little aliasing effect is still present, as clearly visible in the spectra of Figure~\ref{fig:viking_fk}.

\section{Conclusions}\label{sec:conclusion}
In this manuscript, we propose a Deep Prior method for interpolating seismic data strongly affected by aliasing due to high regular sub-sampling.
First, the interpolation problem is cast in the Deep Prior paradigm. This means we make use of a CNN as an optimizer to minimize a cost function, rather than a tool that needs to be trained beforehand.
Then, we add a regularization term that relies on directional Laplacian. The latter is built upon data slopes estimated from a low-pass version of the interpolated data produced in a first step.

Numerical simulations have been run on synthetic, post-stack and pre-stack data. Results have proven the proposed methodology to be effective and worth of future efforts, which will be devoted to improving the robustness of the inversion and extending this paradigm to 3D data.

\bibliographystyle{IEEEbib}
\bibliography{references}

\begin{thebibliography}{10}

\bibitem{chang20063d}
W.~Chang and G.A. Mcmechan,
\newblock ``3d acoustic prestack reverse-time migration,''
\newblock {\em Geophysical Prospecting}, vol. 38, no. 7, pp. 737--755, 2006.

\bibitem{virieux2009overview}
J.~Virieux and S.~Operto,
\newblock ``An overview of full-waveform inversion in exploration geophysics,''
\newblock {\em Geophysics}, vol. 74, no. 6, pp. WCC1--WCC26, 2009.

\bibitem{verschuur1992adaptive}
D.J. Verschuur, A.J. Berkhout, and C.P.A. Wapenaar,
\newblock ``Adaptive surface-related multiple elimination,''
\newblock {\em Geophysics}, vol. 57, no. 9, pp. 1166--1177, 1992.

\bibitem{spitz1991seismic}
S.~Spitz,
\newblock ``Seismic trace interpolation in the fx domain,''
\newblock {\em Geophysics}, vol. 56, no. 6, pp. 785--794, 1991.

\bibitem{huang2018dreamlet}
W.~{Huang}, R.~{Wu}, and R.~{Wang},
\newblock ``Damped dreamlet representation for exploration seismic data
  interpolation and denoising,''
\newblock {\em IEEE Transactions on Geoscience and Remote Sensing (TGRS)}, vol.
  56, no. 6, pp. 3159--3172, 2018.

\bibitem{kumar2017beating}
R.~{Kumar}, O.~{López}, D.~{Davis}, A.~Y. {Aravkin}, and F.~J. {Herrmann},
\newblock ``Beating level-set methods for 5-d seismic data interpolation: A
  primal-dual alternating approach,''
\newblock {\em IEEE Transactions on Computational Imaging}, vol. 3, no. 2, pp.
  264--274, 2017.

\bibitem{lopez2016offthegrid}
O.~{López}, R.~{Kumar}, Ö. {Yılmaz}, and F.~J. {Herrmann},
\newblock ``Off-the-grid low-rank matrix recovery and seismic data
  reconstruction,''
\newblock {\em IEEE Journal of Selected Topics in Signal Processing (JSTSP)},
  vol. 10, no. 4, pp. 658--671, 2016.

\bibitem{nazari2017simultaneous}
M.~A. {Nazari Siahsar}, S.~{Gholtashi}, E.~{Olyaei Torshizi}, W.~{Chen}, and
  Y.~{Chen},
\newblock ``Simultaneous denoising and interpolation of 3-d seismic data via
  damped data-driven optimal singular value shrinkage,''
\newblock {\em IEEE Geoscience and Remote Sensing Letters}, vol. 14, no. 7, pp.
  1086--1090, 2017.

\bibitem{carozzi2021interpolated}
Fernanda Carozzi and Mauricio~D. Sacchi,
\newblock ``Interpolated multichannel singular spectrum analysis: A
  reconstruction method that honors true trace coordinates,''
\newblock {\em Geophysics}, vol. 86, no. 1, pp. A1--V89, 2021.

\bibitem{mandelli2019interpolation}
Sara Mandelli, Federico Borra, Vincenzo Lipari, Paolo Bestagini, Augusto Sarti,
  and Stefano Tubaro,
\newblock ``Seismic data interpolation through convolutional autoencoder,''
\newblock in {\em SEG Technical Program Expanded Abstracts}, pp. 4101--4105.
  2018.

\bibitem{siahkoohi2018seismic}
Ali Siahkoohi, Rajiv Kumar, and F~Herrmann,
\newblock ``Seismic data reconstruction with generative adversarial networks,''
\newblock in {\em 80th EAGE Conference and Exhibition}, 2018.

\bibitem{oliveira2018interpolating}
D.A.B. Oliveira, R.S. Ferreira, R.~Silva, and E.V. Brazil,
\newblock ``Interpolating seismic data with conditional generative adversarial
  networks,''
\newblock {\em IEEE Geoscience and Remote Sensing Letters}, vol. 15, no. 12,
  pp. 1952--1956, 2018.

\bibitem{yoon2020seismic}
D.~{Yoon}, Z.~{Yeeh}, and J.~{Byun},
\newblock ``Seismic data reconstruction using deep bidirectional long
  short-term memory with skip connections,''
\newblock {\em IEEE Geoscience and Remote Sensing Letters (GRSL)}, pp. 1--5,
  2020.

\bibitem{Ulyanov_2018_CVPR}
D.~Ulyanov, A.~Vedaldi, and V.~Lempitsky,
\newblock ``Deep image prior,''
\newblock in {\em The IEEE Conference on Computer Vision and Pattern
  Recognition (CVPR)}, June 2018.

\bibitem{liu2019deep}
Q.~Liu, L.~Fu, and M.~Zhang,
\newblock ``Deep-seismic-prior-based reconstruction of seismic data using
  convolutional neural networks,''
\newblock {\em Geophysics}, 2019.

\bibitem{kong2020eage}
Fantong. Kong, V.~Lipari, P.~Bestagini, and S.~Tubaro,
\newblock ``A deep prior convolutional autoencoder for seismic data
  interpolation,''
\newblock in {\em 82nd EAGE Conference and Exhibition}, 2020.

\bibitem{kong2020seg}
Fantong Kong, Francesco Picetti, Vincenzo Lipari, Paolo Bestagini, and Stefano
  Tubaro,
\newblock ``Deep prior-based seismic data interpolation via multi-res u-net,''
\newblock in {\em Society of Exploration Geophysicists (SEG) Annual Meeting},
  2020.

\bibitem{kong2020grsl}
Fantong Kong, Francesco Picetti, Vincenzo Lipari, Paolo Bestagini, Xiaoming
  Tang, and Stefano Tubaro,
\newblock ``Deep prior-based unsupervised reconstruction of irregularly sampled
  seismic data,''
\newblock 2020.

\bibitem{vanvliet1995estimators}
Lucas~J. van Vliet and Piet~W. Verbeek,
\newblock ``Estimators for orientation and anisotropy in digitized images,''
\newblock in {\em Advanced School for Computing and Imaging Annual Conference
  (ASCI)}, 1995.

\bibitem{hale2007local}
Dave Hale,
\newblock ``Local dip filtering with directional laplacians,''
\newblock {\em CWP Report 657}, 2007.

\bibitem{silva2019netherlands}
Reinaldo~Mozart Silva, Lais Baroni, Rodrigo~S. Ferreira, Daniel Civitarese,
  Daniela Szwarcman, and Emilio~Vital Brazil,
\newblock ``Netherlands dataset: A new public dataset for machine learning in
  seismic interpretation,'' 2019.

\end{thebibliography}

\end{document}